\documentclass{appolb}
\usepackage{epsfig}

\newcommand{\nn}{\nonumber}
\newcommand{\beq} {\begin{equation}}
\newcommand{\eeq} {\end{equation}}
\newcommand{\beqa} {\begin{eqnarray}}
\newcommand{\eeqa} {\end{eqnarray}}

\newcommand{\as}{\alpha_s}

\newcommand{\ieps}{i\varepsilon}

\newcommand{\morder}[1]{{\cal O}\left(#1 \right)}
\newcommand{\eq}[1]{(\ref{#1})}

\newcommand{\ket}[1]{\vert{#1}\rangle}

\newcommand{\ave}[1]{\langle{#1}\rangle}

\newcommand{\pvec}{{\bf p}}

\newcommand{\halft}{{\textstyle \frac{1}{2}}}

\newcommand{\Slash}[1]{ \parbox[b]{0.6em}{$#1$} \hspace{-0.55em}
                                \parbox[b]{0.55em}{ \raisebox{-0.2ex}{$/$}}}

\begin{document}
\title{Filling Perturbative Ground States%
\thanks{Prepared for the Festschrift in honor of Professor Jan Kwieci\'nski. Research supported in part by the European Commission under contract HPRN-CT-2000-00130.}%
}
\author{Paul Hoyer
\address{Department of Physical Sciences and Helsinki Institute of Physics\\
POB 64, FIN-00014 University of Helsinki, Finland}
} 

\maketitle
\begin{abstract}
I discuss a degree of freedom in formulating perturbation theory that is often neglected: the $in$- and $out$-states need not be empty. The inclusion of (free) particles in the asymptotic states modifies the on-shell prescription of the free propagator. This affects loop contributions -- but the modified expansion is {\it a priori} as justified as the standard one with Feynman prescription.

It is possible to dress the quark propagator to all orders with zero-momentum gluons from the perturbative ground state. The dressed quark has no pole and thus cannot appear as an external particle in the S-matrix. Chiral symmetry may be spontaneously broken, but Lorentz and gauge symmetry is exact. Adding loop corrections to this ``dressed tree approximation'' gives a formally exact PQCD expansion. 
\end{abstract}

\vspace{-12cm} \hfill HIP-2003-18/TH \\ \phantom{a} \hspace{\fill} hep-ph/0304022\\
\vspace{11cm}

\section{Introduction}

I have had the honor and pleasure of knowing Jan Kwieci\'nski during most of my research career. Our early work together \cite{Hoyer:hf,Hoyer:1978zq} concerned the analytic structure and high energy behavior of hadron amplitudes. 30 years later, we are still both active in this area. In celebration of his 65th birthday I shall informally discuss a longstanding interest of mine into perturbative QCD with asymptotic ($t\to\pm\infty$) states that contain free quark and gluon pairs. I discussed this line of research publicly for the first time at the 1982 Zakopane school \cite{Hoyer:1982ni}. The recent developments \cite{Hoyer:2002ru,hp} address the analytic structure of quark and gluon amplitudes in the long distance regime.

Like many others, I continue to be tantalized by the fact that we have known Quantum Chromodynamics (QCD) for 30 years and yet are unable to understand (except numerically) how hadrons and color confinement arise. The non-relativistic quark model gives a good phenomenological description of hadrons, making it a candidate for the first term in a systematic approximation scheme of QCD. At the same time we know that hadrons actually are highly relativistic states (98\% of the proton mass is dynamical in origin). The QCD ground state apparently has a condensate of quarks and gluons \cite{svz} which causes a spontaneous breaking of chiral symmetry. How can we derive the simple intuitive quark model using rigorous QCD methods?

My approach has been to have a closer look at perturbation theory. The quark model shows that (constituent) quarks are relevant degrees of freedom. Concepts such as orbital angular momentum that in atomic physics arise through QED perturbation theory are relevant also for the hadron spectrum. Phenomenological analyses of high energy processes indicate that hadron momentum distributions are well described by perturbatively calculated parton distributions down to low momenta \cite{yd}.

Perturbation theory is compelling also because there are few alternatives. Most quantitative results in QED as well as in QCD rely on the perturbative expansion. The series is formally exact and determined by the lagrangian -- well, almost. We need to specify the renormalization scheme and sometimes resum leading logarithms, but this is well understood. My focus is on another choice that we usually make without comment or even recognition: we always expand around an {\em empty} state ($\ a\ket{0} = 0$). Formally, we are allowed to assume empty $in$- and $out$-states at asymptotic times ($t\to\pm\infty$), since the interaction terms in the hamiltonian will turn them into the true ground state during the propagation to finite $t$.

At lowest order of perturbation theory, however, nothing happens between $t = \pm\infty$
and $t=0$. The perturbative expansion starts from the empty $in$- and $out$-states. Given that the QCD ground state is a condensate of partons, it is understandable that standard PQCD fails qualitatively at low momentum scales where interactions with condensate particles are significant. On the other hand, at scales $Q$ which are much higher than that of the condensate ($\Lambda_{QCD} \simeq 200$ MeV) the condensate contributions are power suppressed in $\Lambda_{QCD}/Q$. The failure of the perturbative expansion can be caused by interactions with the condensate even though $\as$ is moderate. It is in fact plausible that the $Q^2$-dependence of $\as$ freezes, $\as(Q^2\to 0) \simeq 0.5$ \cite{yd}.

It thus seems motivated to consider the properties of perturbative expansions around a free state which contains gluon or quark pairs. This is an interesting question regardless of its possible phenomenological relevance. Perturbation theory is such an important tool that we need to explore all its nooks and crannies.

\section{The Boundaries of Path Integrals}

In a path integral derivation of perturbation theory the choice of $in$- and $out$-states is usually swept neatly under the rug. Let me recall the argument for $\phi^4$ theory,
\beq\label{lagrange}
{\cal L} = \frac{1}{2}\left[(\partial_\mu \phi)^2 - m^2\phi^2\right] - \frac{\lambda}{4!}\, \phi^4 \equiv {\cal L}_0 + {\cal L}_{int}
\eeq

We express the generating functional $Z[J]$ of Green functions by extracting the interactions as derivatives wrt. the sources $J$,
\beq \label{zint}
Z[J]= \exp\left[iS_{int}\left(\frac{\delta}{\delta J}\right)\right] Z_0[J]
\eeq
where the free functional $Z_0[J]$ contains only gaussian path integrals which can be done exactly,
\beqa 
Z_0[J] &=& \int {\cal D}[\phi] \exp\left\{i\int d^4x \left[{\cal L}_0(\phi) + J(x)\phi(x)\right]\right\} \label{zfreex} \\
&=& \exp\left[-\frac{1}{2}\int\frac{d^4p}{(2\pi)^4} J(-p) D(p) J(p) \right]
\label{zfreep}
\eeqa
The free propagator is found to be
\beq \label{feynprop}
iD_F(p)=\frac{i}{p^2-m^2+\ieps}
\eeq
and the Feynman $\ieps$ prescription is often motivated by the convergence of the path integral. The perturbative expansion follows by expanding the $\exp(iS_{int})$ factor in \eq{zint}.

This elegant derivation gets the `correct' result without reference to the $in$- and $out$-states at $t=\pm\infty$. A more pedantic approach\footnote{In retrospect, the main (albeit trivial) point I learnt in \cite{Hoyer:1982ni} was the importance of considering path integrals over a finite time interval, thus exposing the wave functions.} would be to start with a path integral defined over a finite time interval $-T \leq t \leq T$,
\beqa \label{zfreet}
Z_0[J] &=& \int_{t=-T}^{T} {\cal D}[\phi(t,\pvec)] \prod_\pvec \exp\left\{\frac{-i}{2}\int_{-T}^T dt\,
\phi^*(t,\pvec)\bigg[\frac{\partial^2}{\partial t^2} \right. + \pvec^2+ m^2 \bigg] \phi(t,\pvec) \nn\\ &+& i\int_{-T}^T dt\, J(t,-\pvec)\phi(t,\pvec)
 -\left. \frac{C(\pvec)}{2}\Big[|\phi(-T,\pvec)|^2+|\phi(T,\pvec)|^2\Big]\right\}
\eeqa
Here the gaussian wave functions at $t=\pm T$ guarantee the convergence of the path integral. The parameter $C(\pvec)$ is readily determined by recalling that the free hamiltonian corresponding to our theory \eq{lagrange},
\beq \label{h0}
H_0(t) = \frac{1}{2}\int\frac{d^3\pvec}{(2\pi)^3}\left[|\pi(t,\pvec)|^2 +
(\pvec^2+m^2) |\phi(t,\pvec)|^2 \right]
\eeq
is a sum (over 3-momentum ${\pvec}$) of uncoupled harmonic oscillator hamiltonians.
The (empty) ground state wave function of the harmonic oscillator is gaussian, with
\beq\label{cfeyn}
C(\pvec)=E_\pvec \equiv \sqrt{\pvec^2+m^2}
\eeq
It is indeed straightforward to show \cite{Hoyer:2002ru} that as $T\to\infty$ the path integral \eq{zfreet} with this value of $C(\pvec)$ gives \eq{zfreep} with the Feynman propagator \eq{feynprop}.

The more explicit expression \eq{zfreet} of the path integral leads naturally to the next question: What if we choose a gaussian wave function in \eq{zfreet} with weight different from \eq{cfeyn}? From
\beq \label{pairs}
\exp\left[-\halft C(\pvec) |\phi|^2 \right] = \sum_n
\frac{(-1)^n}{2^n n!} \left[C(\pvec)
-E_\pvec\right]^n|\phi(t,\pvec)|^{2n} \exp\left[-\halft E_\pvec\,
|\phi(t,\pvec)|^2 \right]
\eeq
we see that this may be interpreted as a coherent superposition of states containing particle pairs. The gaussian integrals can be done with equal ease for any $C$. The result is that the free generating functional still has the form \eq{zfreep}, but a term proportional to
\beq \label{condterm}
i\left[C(\pvec)-E_\pvec\right]\delta(p^0-E_\pvec).
\eeq
is added to the Feynman propagator $D_F(p)$ in \eq{feynprop}. Thus only the on-shell prescription of the free propagator is affected.

In an exact calculation (formally represented by the full perturbative sum) the various choices of $in$- and $out$-states will relax to the true ground state at finite $t$. Thus we may in principle freely choose the value of $C(\pvec)$, and hence the on-shell prescription of the free propagator, for each value of $\pvec$. However, when $\pvec \neq 0$ the asymptotic states are not (even perturbatively) boost invariant. Hence Lorentz invariance is broken order by order in the perturbative expansion. Explicit Lorentz invariance is preserved only for massless particles with zero momentum in the perturbative ground state \cite{cpm}. With this constraint we have a {\em single} parameter $C(\pvec=0)$, which gives a free massless propagator of the form
\beq \label{lamprop}
iD_\lambda(p)=\frac{i}{p^2+\ieps}+\lambda^2 (2\pi)^4 \delta^4(p)
\eeq
where $\lambda$ has the dimension of mass.

Once we have defined the on-shell prescription of the free propagator, the perturbative expansion follows in the usual way through the Taylor expansion of $\exp(iS_{int})$ in \eq{zint}. The standard Feynman rules thus apply with the expression \eq{lamprop} for the free propagator. An analogous conclusion is reached for fermion fields \cite{Hoyer:2002ru}.

Since the relevant modification occurs in the free sector of the theory, the above discussion for a scalar field is applicable also to transverse gluons in QCD. Adding transverse gluons to the asymptotic states should, in particular, preserve gauge invariance. This was checked in \cite{Hoyer:2000ca} through a calculation of the static QCD potential in Feynman and Coulomb gauge. Since the on-shell prescription was modified for a finite range of $|\pvec|$ the equivalence of the two calculations was quite non-trivial. The calculations that I describe in the next section similarly check gauge invariance in a non-trivial way. It would of course be desirable to have a general proof that gauge invariance remains exact.

\section{Dressing the quark propagator with a gluon condensate {\rm \cite{hp}}}

The modified gluon propagator analogous to \eq{lamprop} is (in Feynman gauge),
\beq\label{gpropmod}
i D_{ab}^{\mu\nu}(p) = -\delta_{ab}\, g^{\mu\nu} \left[\frac{i}{p^2+\ieps}
      + \lambda_g^2 \, (2\pi)^4 \, \delta^4(p)\right]
\eeq
The perturbative expansion of a Green function $G$ is then a power series in two parameters, which may be chosen as $\alpha_s$ and  $\alpha_s \lambda_g^2$,
\beq \label{pertexpg}
G=\sum_{\ell=0}^\infty (\as)^\ell \sum_{n=0}^\infty C_{\ell,n}\ (\as \lambda_g^2)^{n}
\eeq
The power $n$ counts how many times we pick the `condensate' term $\propto \delta^4(p)$ in internal gluon propagators, and the power $\ell$ gives the number of times we take the standard part of the gluon propagator.

We recently calculated the full sum over $n$ for $\ell=0$ in the case of the quark propagator, in the limit of a large number of colors $N\to\infty$ \cite{hp}. For this leading condensate contribution there is a $\delta^4(p)$ factor in each loop, so that we are effectively dealing with a tree approximation. Fig.~1 shows typical diagrams that contribute in this approximation. Non-planar diagrams and quark loops do not contribute in the limit of large $N$ with $g^2N$ fixed \cite{thooft}. We also exclude diagrams that involve scattering among the condensate gluons themselves. Such diagrams are proportional to the volume of space-time, $\delta^4(0)$, and we expect them to factorize from physical quantities (but have no proof of this).

\begin{figure}[t]
\begin{center}
\leavevmode
{\epsfxsize=10truecm \epsfbox{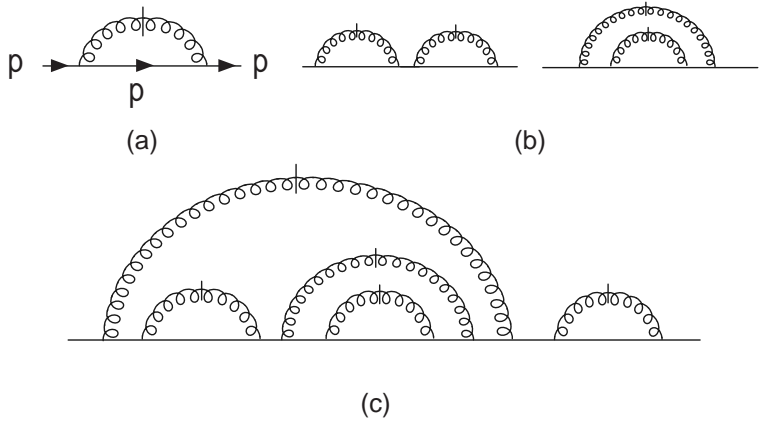}}
\end{center}
\caption[*]{Quark propagator $S_g(p)$ in a gluon condensate. A cut on a gluon line indicates that only the condensate term $\propto \lambda_g^2$ in the free propagator \eq{gpropmod} is taken. (a) Order $\as\lambda_g^2$. (b) Order $\as^2\lambda_g^4$. (c) A generic diagram of order ($\as\lambda_g^2)^{n}$.}
\label{fig1}
\end{figure}

\begin{figure}[b]
\begin{center}
\leavevmode
{\epsfxsize=10truecm \epsfbox{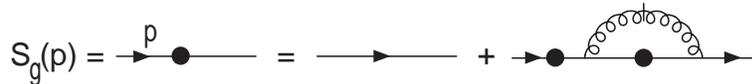}}
\end{center}
\caption[*]{Self-consistency equation for the dressed quark propagator $S_g(p)$. The full blob denotes the gluon condensate dressing.}
\label{fig2}
\end{figure}

The sum over diagrams is most easily done by noting that the self-consistency equation for the dressed quark propagator $S_g(p)$ shown in Fig.~2 generates through iteration all relevant diagrams (\cf Fig.~1). The cut (condensate) gluon carries zero four-momentum. This eliminates the loop integral and reduces this Dyson-Schwinger type equation to an algebraic second order equation for the dressed quark propagator,
\beq
\Slash{p} S_g(p) = 1 - \halft\mu_g^2\, \gamma^{\mu} S_g(p)
\gamma_{\mu} S_g(p)
\label{Sg0}
\eeq
where $C_F \simeq N/2$ in the $N \to\infty$ limit and
\beq
\mu_g^2 = g^2 N \, \lambda_g^2
\label{mug}
\eeq

We find two solutions of \eq{Sg0},
\beqa
S_{g1}(p) &=& \frac{2 \Slash{p}}{p^2 + \sqrt{p^2(p^2-4\mu_g^2)}}\label{csbcons}\\
S_{g2}(p) &=& -\frac{1}{\mu_g^2}\left(\Slash{p} \pm \sqrt{p^2 +
\halft\mu_g^2}\right)\label{csbbreak}
\eeqa
both of which have a novel analytic structure. This is a result of the all-orders sum, and cannot be seen from the first order correction (Fig.~1(a), given by Eq. (2.6) in Ref. \cite{Hoyer:2002ru}).

The $S_{g1}$ solution \eq{csbcons} conserves chiral symmetry and reduces to the free propagator in the $p^2 \to\pm\infty$ limits. But the pole at $p^2=0$ of the free propagator has turned into a $1/\sqrt{p^2}$ branch point in the dressed propagator. This has the consequence that the Fourier transformed propagator
\beq
S_g(t, \pvec) = \int_{-\infty}^{\infty} \frac{dp_0}{2\pi} S_g(p)
\exp(-itp_0)
\eeq
vanishes at large times,
\beq \label{ast}
|S_{g1}(t, \vec{p})| \ \ \mathop{\sim}_{|t| \to \infty} \ \
\morder{1/\sqrt{|t|}}
\eeq
Thus the quark has, startlingly, been removed from the asymptotic states and does not appear in S-matrix elements.

The second solution $S_{g2}$ in \eq{csbbreak} breaks chiral symmetry. It is non-perturbative in the sense that it does not reduce to the free propagator for $\mu_g \to 0$. It also does not approach the free propagator for $p^2 \to\pm\infty$. However, we see that $S_{g1} = S_{g2}$ at $p^2 = -\mu_g^2/2$. Hence it appears possible to pick the $S_{g2}$ solution in a loop integral over the range 
\beq
-\sqrt{\pvec^2-\mu_g^2/2} \leq p^0 \leq \sqrt{\pvec^2-\mu_g^2/2}
\eeq 
This will give amplitudes with spontaneously broken chiral symmetry.

Using the same approximation we evaluated also the dressed quark-photon vertex \cite{hp}.  The Ward identity between the vertex and the inverse propagators is exactly satisfied. The photon self-energy correction leaves the physical photon pole at $p^2=0$. Hence the photon remains an asymptotic, S-matrix state. It would obviously be of interest to study whether this is true more generally for color-singlet states.

\section{Remarks}

Adding a condensate term to the gluon propagator as in \eq{gpropmod} appears to be formally allowed and leads to results which differ in interesting ways from those of standard PQCD in the long distance limit. In particular, the sum of leading condensate contributions allows to define a ``dressed tree approximation'', to which loop corrections ($\ell >0$ in \eq{pertexpg}) can be systematically added. The dressed tree amplitudes have a novel analytic structure as shown by the quark propagators \eq{csbcons}, \eq{csbbreak}. This should allow studies of previous conjectures \cite{Brower:id,Gribov:1999ui} that confined partonic Green functions have an analytic structure which differs from the standard one.

Analyticity is closely related to unitarity. Sums over intermediate states are saturated by hadrons rather than partons in a confining theory. It will thus be interesting to see how unitarity is satisfied in the present framework.

The dressing of the gluon propagator involves a richer set of diagrams than those shown in Fig.~1. The low order contributions ($n=1,2$ in \eq{pertexpg}) have the transverse structure required by gauge invariance \cite{hp,cr}. The expectation value $\ave{F_{\mu\nu}F^{\mu\nu}}$ is non-vanishing due to a contribution from the 4-gluon coupling \cite{cpm,cr}. Hence the study of the gluon sector promises to be rewarding.

\vspace{1cm}

\noindent {\bf Acknowledgements.} This was written in celebration of Jan Kwiecinski, whom I greatly value as a friend and colleague. I thank the organizers of this Festschrift for asking me to write a contribution. Most of the work described above was done in collaboration with St\'ephane Peign\'e.

\end{document}